\documentclass[twoside]{article}
\usepackage{fleqn,espcrc2}

\usepackage{graphicx}
\usepackage[figuresright]{rotating}
\title{SNO and Supernovae}

\author{C.J. Virtue, Laurentian University, Sudbury, Ontario, Canada \\ For the SNO 
Collaboration$^a$} 

\begin{document}

\begin{abstract}
The Sudbury Neutrino Observatory (SNO) has unique capabilities as a supernova detector. In the event
of a galactic supernova there are opportunities, with the data that SNO would collect, to constrain
certain intrinsic neutrino properties significantly, to test details of the various models of supernova dynamics, and to provide
prompt notification to the astronomical community through the Supernova Early Warning System (SNEWS). This paper 
consists of a discussion of these opportunities illustrated by some preliminary Monte Carlo results.
\vspace{1pc}
\end{abstract}

\maketitle

\section{INTRODUCTION}
Supernova neutrinos carry information about both the core collapse process and intrinsic properties of the neutrinos. The Sudbury Neutrino Observatory~\cite{NIM} (SNO) with its capability to differentiate $\nu_e$ from $\overline{\nu}_e$--induced
events, as well as Charged Current (CC) from Neutral Current (NC) events, is well positioned to make unique contributions
to the wealth of physics that would be done with the neutrino signal from the next galactic supernova. 

In the event of a galactic supernova several current generation large-scale detectors would see, over a few tens of seconds, hundreds
to thousands of neutrino-induced events~\cite{SNnu2000}. Many of these detectors have an active program of 
on-line monitoring for such a burst of events and most are presently participating in an inter-experiment collaboration,
the Supernova Early Warning System~\cite{SNEWS1,SNEWS2} (SNEWS). Through this cooperation a supernova could be confirmed,
to a high degree 
of confidence, by the coincident observation of bursts in the sensitive detectors. Such an observed coincidence could
allow a prompt notification of the astronomical community affording them the opportunity to follow the development
of the supernova, with modern instruments, from the earliest possible moment, thus maximizing the scientific return on
the galactic event.

The SNO supernova trigger has been active since May 1999 achieving $\sim 92$\% livetime, a sensitivity that covers $>99$\% of our galaxy, and is now nearing completely automated operation.

\section{NEUTRINOS FROM SUPERNOVAE}

A type II supernova is a prodigious source of neutrinos. The gravitational collapse of an ${\sim1.4}~{\rm M}_{\odot}$ 
stellar core releases approximately $3\times10^{53}$ ergs of gravitational binding energy. A full 99\% of this energy
leaves the proto-neutron star as neutrinos, in a few tens of seconds, roughly evenly distributed across the three flavours
and particle/antiparticle. This represents, in the span of a few seconds, roughly a thousand times more
neutrinos than our sun will produce in its entire $\sim$10 billion year life-time. 

The average energy of supernova neutrinos is of order 15 MeV. Coupled with distance scales of a few kpc, the observation of supernova neutrinos would allow a brief peek at an $L/E$ neutrino oscillation regime not accessible in terrestrial experiments. Neutrino oscillations within the supernova are also interesting to explore with data for multiple neutrino types. Two possibilities exist for a sharp timing signal which would facilitate the extraction of neutrino mass information from the neutrino arrival time spectrum. The first is the ms-scale risetime in neutrino luminosity and the second possibility is the sharp flux cutoff that would accompany the formation of a black hole. Beacom et al.~\cite{BH-cutoff} argue that black hole formation during the period of high neutrino flux is a reasonably probable outcome with tremendous advantages to the direct measurement of neutrino masses. Although supernovae occur at a rate of $\sim$1/sec in the universe at large, the estimated supernova rate in our galaxy seems to range from an optimistic 1/10 years to a more conservative 1/30-50 years~\cite{BH-cutoff,Nature}.
\begin{figure*}[!t]
\includegraphics*[width=16cm]{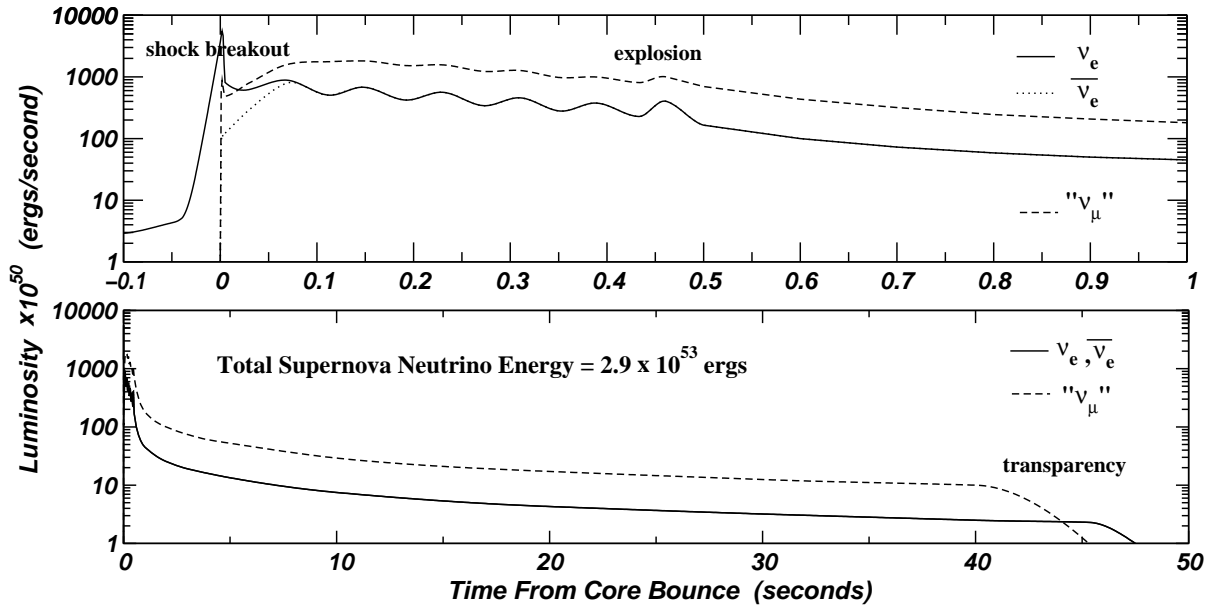}
\caption{Luminosity evolution from the model of Burrows et al.~\protect\cite{Burrows92} used in the SNO supernova Monte Carlo.}
\label{Burlum}
\end{figure*}

The potential of supernova neutrino physics was well demonstrated by the flurry of physics results which followed SN1987A~\cite{SN87A-Gen}. Three detectors, Kamiokande-II~\cite{SN87A-Kam}, IMB~\cite{SN87A-IMB}, and Baksan~\cite{SN87A-Bak} observed a total of 24 neutrino events from the explosion of an ${\sim20}~{\rm M}_{\odot}$ blue supergiant in the LMC. Based on these relatively meager statistics it was possible to set several limits~\cite{SN87A-phys}, of unprecedented precision, particularly on properties of the $\overline{\nu}_e$.

\begin{figure*}[!t]
\includegraphics*[width=16cm]{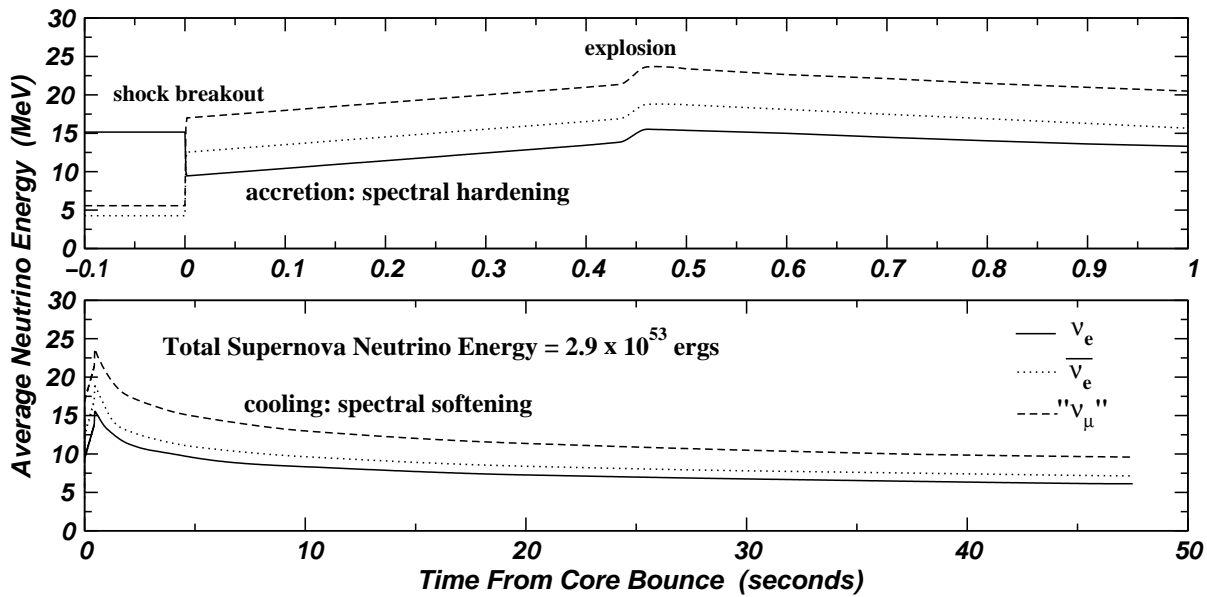}
\caption{Energy evolution from the model of Burrows et al.~\protect\cite{Burrows92} used in the SNO supernova Monte Carlo.}
\label{Burener}
\end{figure*}

Just as neutrinos are our only window into the solar core they are also the only window into the interior of a supernova and the physics of stellar core collapse. Through measurements of the neutrino energy, luminosity and flavour evolution during the course of a supernova it is possible to extract information about the explosion mechanism, proto-neutron star cooling, and black hole formation. Such observations could then be compared with the predictions of current models of these processes. The SNO collaboration's efforts to date to simulate a supernova signal in the detector~\cite{Jaret} have been focused on the 1992 model of Burrows et al.~\cite{Burrows92}. Figure~\ref{Burlum} shows the neutrino luminosity as a function of neutrino species and time that is used as input to the SNO supernova Monte Carlo.
Similarly figure~\ref{Burener} shows the input assumptions for the evolution of the average energy of the different neutrino 
species as a function of time. In both of these figures the first second of time since core bounce is shown on an expanded scale in the upper plot, and ``$\nu_{\mu}$" refers to $\nu_{\mu}$, $\nu_{\tau}$ and their antiparticles, which are indistinguishable in the detector. 

Although this is only one of many models of supernova dynamics this model contains the robust and generic features expected to be present in a supernova neutrino signal. As such it is useful to illustrate the potential sensitivity of SNO to features that would permit a detailed test of supernova models. 

The time regime may be divided into three regions, a collapse phase, prior to core bounce at 0 seconds; an accretion phase, from 0 to 0.45 seconds in this model; and a cooling phase, for times following the explosion at 0.45 seconds. In the collapse phase neutronization produces a sharp burst of $\nu_e$'s. Core bounce is accompanied by an abrupt $\overline{\nu}_e$ and ``$\nu_{\mu}$" turn-on in the high temperature wake of the launched shock wave. Throughout the accretion phase there is gradual spectral hardening of all the neutrino species as the infalling stellar matter heats the early proto-neutron star. The average neutrino energies of the different species are determined by the neutrinosphere radius, for that species, with the smaller radii being hotter. Following a sudden spectral hardening that occurs at the point of explosion, where the flow of infalling matter is reversed, the trend throughout the cooling phase is one of slow spectral softening. 

As discussed in the following section it is the deuterium in the heavy water and the arising distinct event signatures that give SNO the capability to distinguish between various neutrino species in the supernova neutrino signal. Figures~\ref{Burlum} and \ref{Burener} therefore illustrate well the rich possibilities for constraining supernova models with data in the case where sufficient statistics are available and these unique capabilities are exploited. By contrast a full 95\% of the supernova events in a light water Cerenkov detector are attributable to the $\overline{\nu}_e$ flux alone.

\section{SNO AS A SUPERNOVA DETECTOR}

The Sudbury Neutrino Observatory detector and its early results are described elsewhere in these proceedings by Noble~\cite{TonyNOW}. As a supernova detector SNO consists of 1000 tonnes of heavy water and about 1400 of light water efficiently viewed by over 9000 PMTs. Both the heavy and light water are active detector volumes for supernova detection. The detector is located 2km underground, at 6000 mwe depth, in a carefully controlled and radioactively very quiet environment. The high-energy background at this depth is due to cosmic ray muons at a rate of about 3 hr$^{-1}$ with a smaller, low-energy contribution from solar neutrino-induced reactions. Sensitivity to galactic supernova is efficiently achieved by the simple trigger condition of 50 events, above an approximately 4 MeV threshold, within a 2 second window. Some on-line processing of the data stream is required to deal with an occasional instrumental event burst and to remain ``live" to supernovae through most detector calibration procedures. The SNO front-end electronics has the capability to accept burst data rates approaching 2 MHz for short times, and has the capacity to buffer approximately $10^6$ typical neutrino events for later readout. 

\begin{figure}[!t]
\vspace*{0.3cm}
\includegraphics*[width=7.5cm]{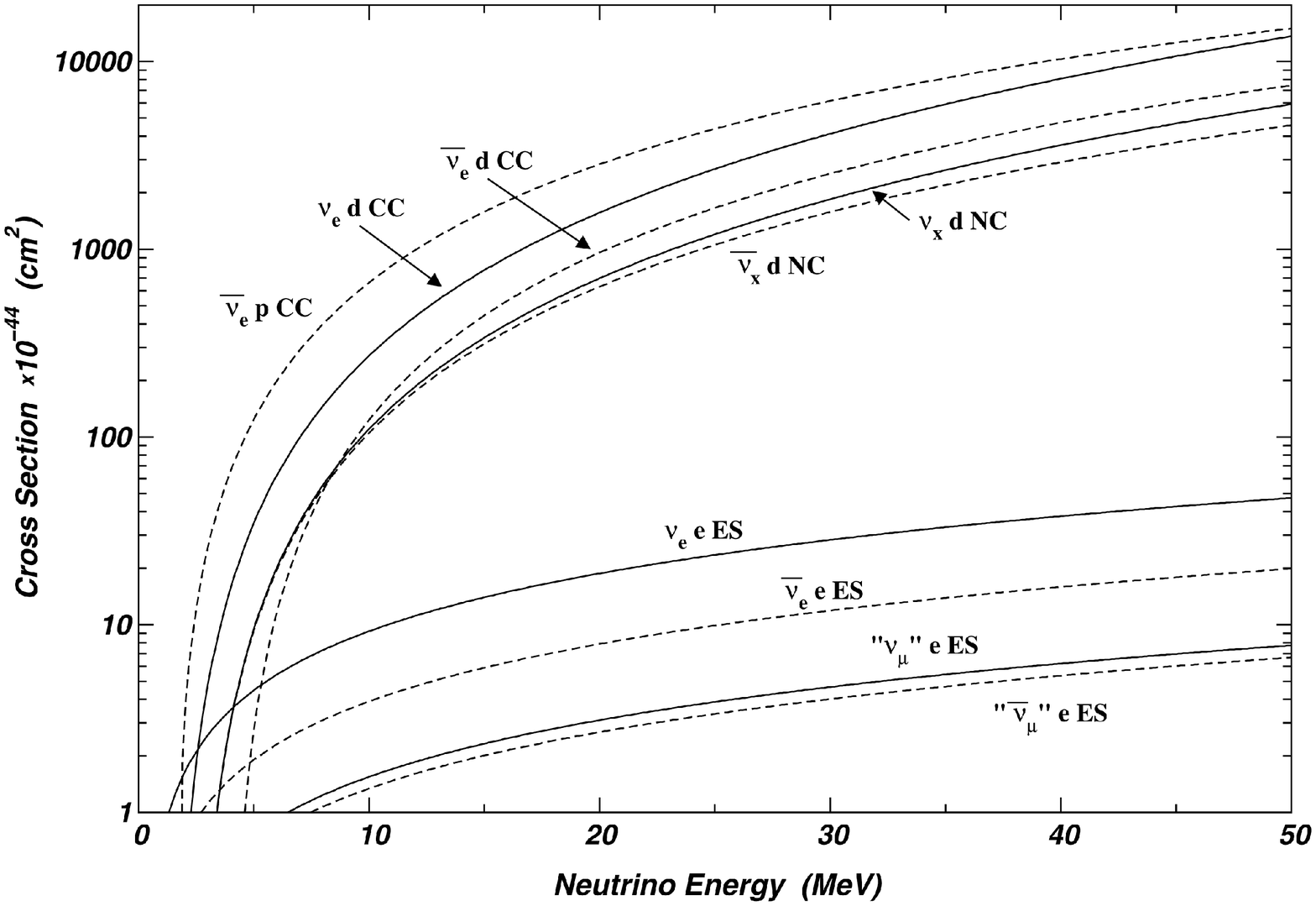}
\caption{Supernova neutrino cross-sections relevant to the SNO detector.}
\label{SNOxsect}
\end{figure}

Figure~\ref{SNOxsect} shows the cross-sections as a function of neutrino energy for the charge current (CC), neutral current (NC), and elastic scattering (ES) reactions in heavy and light water. The combination of the NC and CC reactions on deuterium roughly matches the $\overline{\nu}_e$ CC cross-section on protons in the light water resulting in the statistics for a supernova being divided approximately equally between the two water volumes. As seen in the reactions listed in table~\ref{SNMCsalt}, the advantage of the heavy water is in the possibility to distinguish between the $\nu_e$ CC reaction, with a single energetic electron; the $\overline{\nu}_e$ CC reaction, with an energetic positron and two neutron captures; and the combined NC reactions, with no electron and a single neutron capture signal. 
\begin{table*}[!t]
\caption{Breakdown of the Supernova Monte Carlo Results at 10 kpc, for the ``Salt" Running Phase}
\label{SNMCsalt}
\newcommand{\m}{\hphantom{$-$}}
\newcommand{\cc}[1]{\multicolumn{1}{c}{#1}}
\renewcommand{\tabcolsep}{1.8pc} 
\renewcommand{\arraystretch}{1.2} 
\begin{tabular*}{16cm}{lcccc}
\multicolumn{5}{l}{In 1k tonne of heavy water} \\
\hline
Reaction                 & \cc{\#Events} & \cc{Energy} & \cc{Time} & \cc{Pointing} \\
\hline
CC:~$\nu_e+d\rightarrow p+p+e^-$                           & 72  & y      & y       & $\ast$  \\
CC:~$\overline{\nu}_e+d\rightarrow n+n+e^+$                & 138 & y      & y       & $\ast$ \\
NC:~$\nu_e+d\rightarrow \nu_e+p+n$                         & 30  & n      & $\sim$y &  n \\
NC:~$\overline{\nu}_e+d\rightarrow \overline{\nu}_e+p+n$   & 32  & n      & $\sim$y &  n \\
NC:~$\nu_{\mu , \tau}+d\rightarrow \nu_{\mu , \tau}+p+n$   & 164 & n      & $\sim$y &  n \\
ES:~$\nu_e+e^-\rightarrow\nu_e+e^-$                        & 8   & $\sim$ & y       &  y \\
ES:~$\overline{\nu}_e+e^-\rightarrow \overline{\nu}_e+e^-$ & 3   & $\sim$ & y       &  y \\
ES:~$\nu_{\mu , \tau}+e^-\rightarrow \nu_{\mu , \tau}+e^-$ & 4   & $\sim$ & y       &  y \\
& & & & \\
\multicolumn{5}{l}{In 1.4k tonnes of light water} \\
\hline
Reaction                 & \cc{\#Events} & \cc{Energy} & \cc{Time} & \cc{Pointing} \\
\hline
CC:~$\overline{\nu}_e+p\rightarrow n+e^+$                  & 331& y      & y & $\ast$ \\
ES:~$\nu_e+e^-\rightarrow\nu_e+e^-$                        & 12 & $\sim$ & y & y \\
ES:~$\overline{\nu}_e+e^-\rightarrow \overline{\nu}_e+e^-$ & 5  & $\sim$ & y & y \\
ES:~$\nu_{\mu , \tau}+e^-\rightarrow \nu_{\mu , \tau}+e^-$ & 5  & $\sim$ & y & y \\
\hline
\end{tabular*}\\[2pt]
Note: $\nu_{\mu , \tau}=``\nu_{\mu}"=\nu_{\mu}+\overline{\nu}_{\mu}+\nu_{\tau}+\overline{\nu}_{\tau}$ and the significance of $\ast$ and  $\sim$ is discussed in the text.
\end{table*}

Neutron detection in SNO is as essential to the analysis of a supernova neutrino burst as it is to solar neutrino data. To meet the objectives of the solar neutrino measurement program SNO is foreseen to run in three distinct modes with different neutron detection methods and characteristics. In its first running phase, with pure ${\rm D_2O}$, neutrons are detectable by capture on deuterium to form tritium accompanied by a 6.3 MeV gamma-ray. The capture efficiency is only 24\% and the capture is indistinguishable from CC events, on an event-by-event basis, except to the extent that the signal is seen at lower energies than the typical supernova neutrino CC events. The neutron capture life-time in pure ${\rm D_2O}$ is 40 ms. In order to enhance the neutron detection efficiency ${\rm ^{35}Cl}$ will be added to the ${\rm D_2O}$ in the form of NaCl for the second ``salt" running phase. This will raise the neutron capture efficiency to approximately 83\% and shorten the neutron capture life-time to 4 ms. ${\rm ^{36}Cl}$ de-excites via an 8.6 MeV total energy gamma-ray cascade which is statistically distinguishable from CC (single energetic particle Cerenkov) events by angular isotropy measures. As a third and longer term running configuration the salt added in the second phase will be removed and SNO will install $\sim700$ m of ${\rm ^3He}$ proportional counters on a 1 m by 1 m grid throughout the ${\rm D_2O}$ volume. Neutron capture on ${\rm ^3He}$ will provide positive NC identification on an event-by-event basis with a capture efficiency of 45\%. An additional 12\% will still capture on ${\rm D_2O}$ with a combined neutron capture life-time of 16 ms. For a supernova at 10 kpc the Monte Carlo program with realistic thresholds and efficiencies gives 606, 804, and 681 detected events in pure ${\rm D_2O}$, with added salt, and with the ${\rm ^3He}$ ``Neutral Current Detectors" (NCD) respectively.

Table~\ref{SNMCsalt} summarizes the number of events, for a 10 kpc distant supernova, broken down by the reaction in either the light or heavy water. Also summarized is the quality of the information, provided by each reaction channel, for the energy, interaction time, and direction of the supernova neutrinos. For elastic scattering channels the energy information is comparatively poor but timing and directional information are good. Even the small number of ES events, for a 10 kpc supernova, will be adequate to determine the stellar coordinates of the supernova with an approximate $25^{\circ}$ resolution. In the case of the neutral current channels the timing information is slightly degraded by the neutron capture times and no useful energy or directional information is possible from the data. The charged current channels provide good energy and timing information. However the angular asymmetries of the electron wrt the neutrino direction are weak, energy dependent, and change sign~\cite{angdist} in going from light to heavy water (i.e. as a function of radius). The net effect is that they sum to an approximately isotropic distribution from which it is challenging, if not impossible, to extract directional or ``pointing" information even with sophisticated effort. 

\section{SNEWS}

Depending on the mass and nature of the progenitor star neutrinos decouple and escape from the supernova anywhere from 30 minutes to 10 hours before the first photons. The opportunity is therefore present, if the neutrino experiments are able to provide accurate directional information, to alert the astronomical community to the next galactic supernova and enable them to observe it from the earliest possible times. Without directional information the earliest times are very likely to go unobserved and some part of the unique opportunity is lost. In the early stages of a supernova one may learn about the progenitor and its environment. A UV / soft X-ray flash is expected at shock breakout that would illuminate the surroundings.

The inter-experiment Supernova Early Warning System (SNEWS) aims to provide a fast and very reliable alert to the astronomical community by forming a coincidence of burst signals from several operating detectors. Individual detectors employ a burst detection algorithm with a threshold adjusted such that they are highly efficient for galactic supernova and have an false alert rate of not more than one per week on average. The individual alert signals are transmitted as secure encrypted datagrams via the Internet to redundant coincidence servers located in Japan and Italy. An alert to the astronomical community would be issued if there was a coincidence within a 10 second window in the UT timestamped datagrams from two or more of the participating experiments. If the individual alerts are poissonian then the false coincidence rate is of order once per century. More details on the current status and operation of SNEWS may be found in references~\cite{SNnu2000,SNEWS1,SNEWS2}. SNO is preparing to provide a signal for SNEWS in the near future.

In the case of the SNO supernova trigger the threshold has been set such that alerts would be forwarded to the coincidence servers only when the event burst is identified by in-line analysis to be largely consistent with Cerenkov light in the detector. At present the only regular occurences which satisfy this criteria are muon spallation events followed by multiple pion decays and neutron captures. Such events occur in the SNO detector, because of the great depth, roughly every second month.

The SNEWS objective of providing a fast and reliable alert to the astronomical community is ensured by the redundancy of the coincidence requirement. Our high standards for a low false alarm rate should carry over into efforts to obtain redundant directional information if at all possible as the consequence of an incorrect or misleading single determination is likely to be a significant loss of opportunity. Extreme care is being devoted to avoiding any false alarms. Discussions are continuing among the SNEWS members to decide on the information that can be provided promptly in the event of a supernova, particularly directional information. 

\section{CONCLUSIONS}
The SNO detector's capabilities to distinguish between $\nu_e$, $\overline{\nu}_e$ and NC interactions are unique among existing detectors and, in the event of a galactic supernova, represent an extraordinary opportunity to probe and constrain supernova dynamics and neutrino physical properties. Our participation in SNEWS should help to provide a reliable early warning to the astronomical community of a nearby supernova. Work is continuing on maximizing the detector livetime for supernova and on ensuring the reliability of the on-line supernova trigger under all detector operating conditions. Where statistics permit redundant determinations of the supernova direction would be highly desirable.

\section{ACKNOWLEDGMENTS}

It is a pleasure to acknowledge the contributions of Peter Doe, Jaret Heise, Mike Schwendener and R\'{e}da Tafirout as the other members of SNO's Supernova Trigger Group.

$^a$The SNO Collaboration includes: M. G. Boulay, E. Bonvin, M. Chen, F. A. 
Duncan, E. D. Earle, H. C. Evans, G.T. Ewan, R. J. Ford, A. L. Hallin, P. J. 
Harvey, J. D. Hepburn, C. Jillings, H. W. Lee, J. R. Leslie, H. B. Mak, A. B. 
McDonald, W. McLatchie, B. A. Moffat, B.C. Robertson, P. Skensved, B. Sur, {\bf  Queen's 
University, Kingston, Ontario K7L 3N6, Canada.}
% ...

\end{document}